\documentclass[preprint,12pt]{elsarticle}

\usepackage[right]{lineno}
\usepackage[doublespacing]{setspace}
\usepackage{amssymb}
\usepackage{amsthm}
\usepackage[version=4]{mhchem}
\usepackage{lscape}
\usepackage{hyperref}
\hypersetup{colorlinks=true}

\journal{Earth and Planetary Science Letters}

\begin{document}

\begin{frontmatter}



\title{Can spinodal decomposition occur during decompression-induced vesiculation of magma?}

\author[inst1]{Mizuki Nishiwaki}

\affiliation[inst1]{organization={Center for Glass Science \& Technology, School of Engineering, The University of Shiga Prefecture},
            addressline={2500, Hassaka-cho}, 
            city={Hikone, Shiga},
            postcode={522-8533}, 
            country={Japan}}

\begin{abstract}
Volcanic eruptions are driven by decompression-induced vesiculation of supersaturated volatile components in magma. The initial phase of this phenomenon has long been described as a process of nucleation and growth. Recently, it was proposed that spinodal decomposition (an energetically spontaneous phase separation that does not require the formation of a distinct interface) may occur during decompression-induced magma vesiculation. This suggestion has attracted considerable attention, but is currently only based on textural observations of decompression experiment products (the independence of bubble number density on decompression rate; the homogeneous spatial distribution of the bubbles). 
In this study, I used a simple thermodynamic approach to investigate whether spinodal decomposition can occur during the decompression-induced vesiculation of magma. I drew the binodal and spinodal curves on the chemical composition--pressure plane by approximating hydrous magmas at several conditions of temperature and chemical composition as two-component symmetric regular solutions of silicate and water, and using experimentally determined values of water solubility in these magmas. The spinodal curve was much lower than the binodal curve for all the magmas at pressures sufficiently below the second critical endpoint. In addition, the final pressure of all the decompression experiments performed to date fell between these two curves. This suggests that spinodal decomposition is unlikely to occur in the pressure range of magmatic processes in the continental crust, and that decompression-induced magma vesiculation results from nucleation and subsequent growth, as previously considered. In addition, it is expected that by substituting the determined spinodal pressure into the formula of non-classical nucleation theory, the surface tension between the silicate melt and bubble nucleus can be easily estimated. 
\end{abstract}

\begin{highlights}
\item Magma vesiculation during decompression was interpreted by simple thermodynamics.
\item Binodal and spinodal curves of silicate--water systems were drawn versus pressure.
\item Magma vesiculation probably occurs by nucleation, not by spinodal decomposition.
\item Spinodal pressure can contribute to an easy estimate of bubble surface tension.
\end{highlights}

\begin{keyword} 
hydrous magma \sep vesiculation \sep nucleation \sep spinodal decomposition \sep thermodynamics \sep regular solution
\end{keyword}

\end{frontmatter}

\begin{linenumbers}

\begin{table}[h]
\caption{Notation list.}
\begin{flushleft}
\begin{tabular}{lll} \hline
Symbol & Unit & Definition \\ \hline
$a_0$ & m & Average distance between water molecules in the melt \\ 
$c$ & no unit & Water solubility in the melt (mole fraction) \\
$D_\mathrm{\ce{H2O}}$ & m$^2$ s$^{-1}$ & Diffusivity of total water in the melt \\ 
$g^\mathrm{excess}$ & J mol$^{-1}$ & Molar excess Gibbs energy for a regular solution \\
$g^\mathrm{ideal}$ & J mol$^{-1}$ & Molar Gibbs energy of an ideal solution \\
$g^\mathrm{real}$ & J mol$^{-1}$ & Molar Gibbs energy of mixing ($= g^\mathrm{ideal} + g^\mathrm{excess}$) \\
$h^\mathrm{excess}$ & J mol$^{-1}$ & Molar excess enthalpy for a non-ideal solution \\
$J$ & No m$^{-3}$ s$^{-1}$ & Nucleation rate \\ 
$k_\mathrm{B}$ & J K$^{-1}$ & Boltzman's constant \\ 
$n_0$ & No m$^{-3}$ & Number of water molecules per unit melt volume \\ 
$P$ & Pa & Pressure \\
$P_\mathrm{bi}$ & Pa & Pressure on the binodal curve (= binodal pressure) \\
$P_\mathrm{spi}$ & Pa & Pressure on the spinodal curve (= spinodal pressure) \\
$P_\mathrm{M}$ & Pa & Melt pressure \\
$P_\mathrm{SAT}$ & Pa & Water saturation pressure \\
$P^*_\mathrm{B}$ & Pa & Internal pressure of the critical bubble nucleus \\ 
$R$ & J K$^{-1}$ mol$^{-1}$ & Gas constant \\
$s^\mathrm{excess}$ & J K$^{-1}$ mol$^{-1}$ & 
\begin{tabular}{l} 
Molar excess entropy for a non-ideal solution \\ (= 0 for a regular solution) \\
\end{tabular} \\
$T$ & K & Temperature \\
$\overline{V}_\mathrm{\ce{H2O}}$ & m$^3$ mol$^{-1}$ & Partial molar volume of water in the melt \\ 
$w_\mathrm{sym}$ & J mol$^{-1}$ & 
\begin{tabular}{l} 
Interaction parameter between two components \\ for a symmetirc regular solution \\
\end{tabular} \\
$w_\mathrm{A, \,B}$ & J mol$^{-1}$ & 
\begin{tabular}{l} 
Interaction parameter between two components \\ for an asymmetric regular solution \\
\end{tabular} \\
$w_\mathrm{ijk}$ & J mol$^{-1}$ & 
\begin{tabular}{l} 
Interaction parameter between three components \\ for an asymmetirc regular solution \\
\end{tabular} \\
$X$ & no unit & Symbolic notation for chemical composition \\
$x$ & no unit & Mole fraction of one of the two components \\
$x_\mathrm{bi} (P)$ & no unit & The $x$ that constitutes the binodal curve at pressure $P$ \\
$x_\mathrm{spi} (P)$ & no unit & The $x$ that constitutes the spinodal curve at pressure $P$ \\
$\sigma$ & N m$^{-1}$ & 
\begin{tabular}{l} 
``Microscopic'' surface tension between the melt and \\ homogeneous spherical bubble nucleus with a large curvature
\end{tabular} \\
$\sigma_\infty$ & N m$^{-1}$ & 
\begin{tabular}{l} 
``Macroscopic'' surface tension at the flat interface between \\ the melt and vapor 
\end{tabular} \\ \hline
\end{tabular}
\end{flushleft}
\label{Table1}
\end{table}

\section{Introduction}
\label{Introduction}
Magma degassing is one of the strongest controlling factors in the dynamics of volcanic eruptions. 
Volatiles (e.g., \ce{H2O}, \ce{CO2}, \ce{H2S}) initially dissolved in magma deep underground become insoluble under decompression, precipitating as vapor (bubbles). Understanding this magma vesiculation process is crucial because it dramatically influences eruption style and volcanic explosivity, playing a key role in determining the time evolution of eruptions, ranging from effusive lava flows to highly explosive pyroclastic events. In particular, water often makes up a large proportion of the various volatile components, and its degassing---in a precise sense, the phase separation into silicate melt saturated with water and supercritical water vapor saturated with trace amounts of silicate---has been extensively studied since Verhoogen (1951).
The initial degassing stage has long been understood as nucleation and growth (e.g., Shimozuru et al., 1957; Murase and McBirney, 1973; Sparks, 1978), and theoretical numerical models were constructed by Toramaru (1989, 1995) to predict the bubble number density (BND) based on classical nucleation theory (CNT, e.g., Hirth et al., 1970). In addition, in the last 30 years since the innovative work by Hurwitz and Navon (1994), many experiments have been conducted to reproduce decompression-induced vesiculation of mainly water-dissolved (hydrous) magmas in laboratory experimental magma analogues with controlled decompression rates. 
As data from decompression experiments to date generally agreed with numerical predictions (BND $\propto |\mathrm{decompression\ rate}|^{1.5}$ by Toramaru, 1995) based on CNT, Toramaru (2006) constructed the BND decompression rate meter (BND decompression rate meter). 
This equation is approximately valid for both homogeneous nucleation of spherical bubbles in a uniform melt (where $\sigma$ is large) and heterogeneous nucleation of bubbles on crystal surfaces such as Fe--Ti oxides (where $\sigma$ is small) by applying appropriate corrections to the surface tension $\sigma$ between the melt and bubble nucleus (Shea, 2017; Toramaru, 2022).
This theoretical model has been used extensively to estimate magma's decompression rate in a volcanic conduit based on quantitative analysis of bubble texture in natural pyroclastic products (e.g., Toramaru, 2006; Giachetti et al., 2010; Houghton et al., 2010; Nguyen et al., 2014).

However, some laboratory experiments performed thus far have reported results that are anharmonic to the above equations; Allabar and Nowak (2018) performed decompression experiments on hydrous phonolitic melts over a wide range of decompression rates, including 0.024--1.7 MPa/s, and found systematically high BND values (5.2 mm$^{-3}$) independent of decompression rate. To explain this result, they proposed a scenario in which spinodal decomposition, rather than nucleation as used in previous explanations, occurs in the early stages of decompression-induced vesiculation. Spinodal decomposition is the phase separation of a multi-component mixture or solid solution due to energetic instability (e.g., Cahn and Hilliard, 1959; Cahn, 1965). 
When the system's temperature, pressure, and chemical composition $(T, P, X)$ are within the miscibility gap, whether nucleation or spinodal decomposition occurs is determined by the sign of the second-order derivative of the system's molar Gibbs energy of mixing $g^{\mathrm{real}}$.
If the sign is positive, the system is metastable, and nucleation occurs, with distinct phase boundaries (interfaces) appearing spatially random from the beginning. 
Conversely, if the sign is negative, the system becomes unstable, and spinodal decomposition occurs, in which initially small concentration fluctuations with an unclear phase boundary gradually grow and eventually lead to the separation of two phases at a specific wavelength, forming a distinct interface. 
The diffusion coefficient is proportional to the second-order derivative of $g^{\mathrm{real}}$, so nucleation corresponds to downhill diffusion, where diffusion progresses in the direction that weakens the concentration gradient. In contrast, spinodal decomposition corresponds to uphill diffusion, where diffusion progresses in the direction that strengthens the concentration gradient (Haasen, 1996). 
In Allabar and Nowak (2018), spinodal decomposition was proposed for two reasons: (1) the timescale for spinodal decomposition in gas--liquid systems is much shorter than the timescale for decompression (Debenedetti, 2000), which could explain the absence of dependence on BND on the decompression rate. (2) The bubble spatial distribution in the experimental products was homogeneous; the vitrified silicate melt and bubbles appeared to phase-separate at a specific wavelength. 
Subsequent work by Sahagian and Carley (2020) raised the problem that ``the surface tension between the melt and tiny bubble nucleus should act to push dissolved volatiles back into the melt, but bubbles of such size are still formed'' and discussed this process as the ``tiny bubble paradox.'' They extended the ideas of Allabar and Nowak (2018) as follows: if spinodal decomposition---rather than nucleation---occurs, this paradox can be resolved because interface formation is no longer necessary and explains the homogeneous spatial distribution of bubbles observed in some laboratory products. 

Thus, the new theory that ``decompression-induced vesiculation of magma can occur not only by nucleation but also by spinodal decomposition'' has been actively discussed and has attracted much attention in the last seven years. Gardner et al. (2023) also stated that future interpretations of BND and bubble size distributions of natural volcanic products must consider the possibility that various mechanisms of bubble formation may occur, including nucleation (homogeneous and heterogeneous) and spinodal decomposition. However, whether spinodal decomposition actually occurs during decompression-induced magma vesiculation remains speculative. Nonetheless, owing to the small spatiotemporal scale of the physical phenomena under investigation, it is likely to be extremely difficult to confirm via laboratory observational experiments.
Therefore, in this paper, I discuss which nucleation or spinodal decomposition mechanism is likely to occur during decompression-induced magma vesiculation, based on a simple thermodynamic approach.

This study first reviews the thermodynamic definitions of nucleation and spinodal decomposition. Spinodal decomposition, which has traditionally been treated when it occurs with a change in temperature at constant pressure, can also be treated with a change in pressure at constant temperature. An attempt is made to consider hydrous magmas in a simplified way as a symmetric regular solution of silicate and water and to draw binodal and spinodal curves quantitatively on the chemical composition--pressure plane using a simple calculation. 
Next, based on the calculation results, I discussed the possibility of spinodal decomposition occurring in decompressing magma. Furthermore, I provided insights on the kinetic effects that should be considered in real systems and how the results of previous decompression experiments should be interpreted about the BND decompression rate meter. Finally, as a potential application of the model proposed in this study, an estimate of the surface tension between the melt and the bubble nucleus was presented.

\section{Energetics on the mixing of silicate and water}
\label{What}

\subsection{General theory: Thermodynamic energetics of two-component mixture}
\label{Energetics}
\begin{figure}
    \centering
    \includegraphics[width=0.8\linewidth]{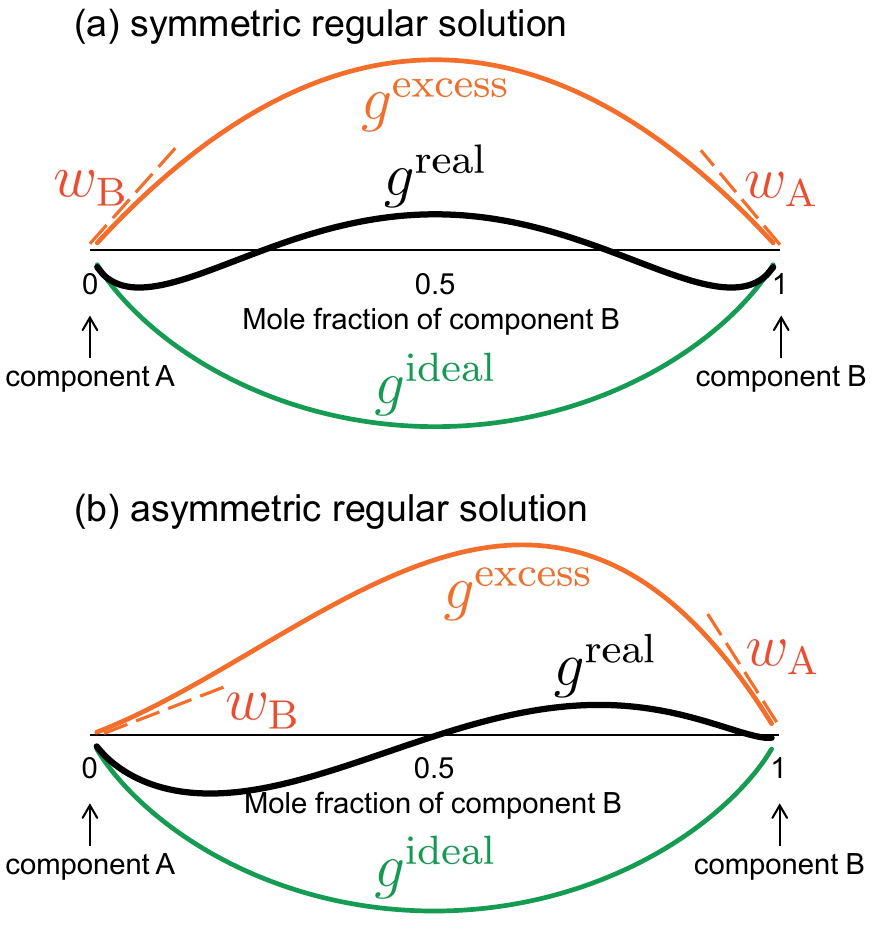}
    \caption{Schematic diagrams of the molar Gibbs energy for regular solutions of two components A and B: (a) symmetric regular solution and (b) asymmetric regular solution.
The green curve represents that of an ideal solution $g^\mathrm{ideal}$, which is common to both regular solution models. 
The orange curve represents the molar excess Gibbs energy for a regular solution $g^\mathrm{excess}$, and the two models are differentiated based on its shape. 
The bold black curve represents the sum of these: the molar Gibbs energy of mixing $g^\mathrm{real}$. 
The interaction parameters $w_\mathrm{A}$ and $w_\mathrm{B}$ are represented by the slopes of the $g^\mathrm{excess}$ curve at the endpoints on the B side and A side, respectively. In the symmetric model (a), $w_\mathrm{A} = w_\mathrm{B}$, which is referred to as $w_\mathrm{sym}$ in the text. 
}
    \label{Fig1}
\end{figure}

Notations used in this study are listed in Table \ref{Table1}. 
Here, I will explain the thermodynamics of a two-component mixture.
In general, the molar Gibbs energy of a mixture, $g^\mathrm{real}$, is described as the sum of the ideal solution's Gibbs energy, $g^\mathrm{ideal}$ (which arises from configurational entropy), and the excess energy, $g^\mathrm{excess}$ ($\neq 0$, represents the deviation from the ideal solution) (e.g., Guggenheim, 1952):
\begin{equation}
\label{g-real}
g^{\mathrm{real}} = g^{\mathrm{ideal}} + g^{\mathrm{excess}},
\end{equation}
where the amount obtained by proportionally distributing and summing the Gibbs energies of each pure phase is subtracted as the baseline. 
In an ideal solution, $g^\mathrm{excess} = 0$. 
Several types of non-ideal solutions that take into account the deviation from the ideal solution, $g^\mathrm{excess} \neq 0$; among them, the most basic type is the regular solution. 
In a regular solution, the non-ideal entropy is neglected ($s^\mathrm{excess} = 0$), and the non-ideal enthalpy is treated as $h^\mathrm{excess} \neq 0$, which leads to the relationship $g^{\mathrm{excess}} = h^{\mathrm{excess}}$.
Here, $x$ represents the mole fraction of one of the components ($0 < x < 1$).
In this case, as explained below, $g^\mathrm{ideal}$ is always symmetric with respect to $x = 0.5$, while the shape of $g^\mathrm{excess}$ (and $g^\mathrm{real}$) is assumed to be either symmetric or asymmetric. The $g^\mathrm{ideal}$ at a given temperature $T$ is expressed by the following equation.
\begin{equation}
\label{g-ideal}
g^{\mathrm{ideal}} = RT \{ x \ln x + (1-x) \ln (1-x) \}.
\end{equation}
The equation for $g^\mathrm{excess}$ differs depending on whether the symmetric or asymmetric model.
In the case of the symmetric model, 
\begin{equation}
\label{sym}
g^{\mathrm{excess}} = x (1-x) w_\mathrm{sym},
\end{equation}
where $w_\mathrm{sym}$ is the interaction parameter between the two components A and B, representing the non-ideality of mixing. 
In the case of the asymmetric model, 
\begin{equation}
\label{asym}
g^{\mathrm{excess}} = x (1-x) \{ w_\mathrm{A} (1-x) + w_\mathrm{B} x \},
\end{equation}
where $w_\mathrm{A}$ and $w_\mathrm{B}$ are the interaction parameters when a particle of A enters a group of B, and when a particle of B enters a group of A, respectively. 
These represent the slopes of the $g^{\mathrm{excess}}$ curve at the endpoints on the B side and A side, where a positive slope indicates that the system becomes energetically unstable after mixing. In contrast, a negative slope indicates that the system becomes energetically stable. 
Based on this, the Gibbs energy curves for both the symmetric and asymmetric models are illustrated as shown in Fig. \ref{Fig1}. 
Here, we assume that for the symmetric model, $w_\mathrm{sym} > 0$, and for the asymmetric model, $w_\mathrm{A} > 0$ and $w_\mathrm{B} > 0$.

\begin{figure}
\centering
\includegraphics{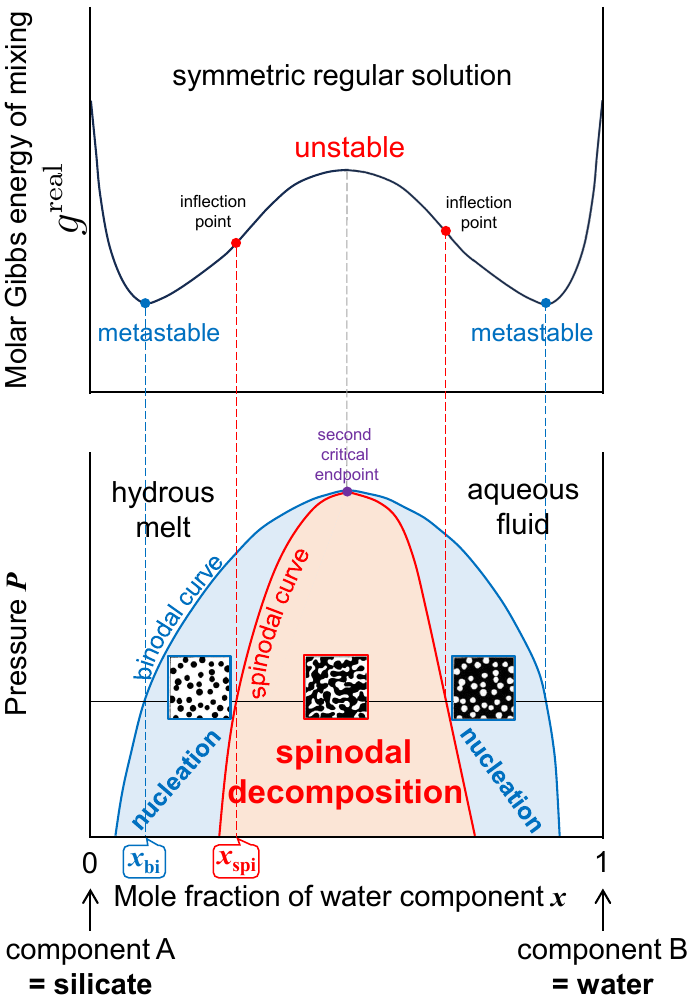}
\caption{Schematic molar Gibbs energy of mixing $g^\mathrm{real}$ (upper panel) and corresponding phase diagram (lower panel) for a general two-component symmetric regular solution. In the interval where the $g^\mathrm{real}$ curve is convex downwards, the system is metastable, and nucleation occurs with clear phase boundaries (surfaces) appearing randomly in space. On the other hand, when the $g^\mathrm{real}$ curve is convex upward, the system is unstable, and spinodal decomposition occurs, in which the two phases start to separate at a specific wavelength with unclear phase boundaries. 
}
\label{Fig2}
\end{figure}

The $g^{\mathrm{real}}$ curve for the symmetric model and its corresponding phase diagram are shown in Fig. \ref{Fig2}. Hereafter, we focus on conditions under which the system’s temperature, pressure, and chemical composition $(T, P, X)$ lie within the miscibility gap. Compositions outside the two local minima of the $g^{\mathrm{real}}$ curve correspond to a single-phase system, while compositions inside these minima undergo phase separation into a two-phase system. In other words, the point where the first derivative of the $g^{\mathrm{real}}$ curve with respect to $x$ equals zero corresponds to the one-phase/two-phase boundary in the phase diagram (i.e., the binodal curve).
In the region between the local minimum and the inflection point, where the $g^{\mathrm{real}}$ curve is convex downwards (i.e., where the second-order derivative with respect to $x$ is positive), the system is metastable, and nucleation occurs with distinct phase boundaries (interfaces) appearing randomly in space. 
Conversely, in the interval where the $g^{\mathrm{real}}$ curve is convex upward, i.e., where the sign of the second-order derivative with $x$ is negative, the system is unstable, and spinodal decomposition occurs wherein the two phases start to separate at a specific wavelength with unclear phase boundaries. 
In other words, the point where the second derivative of the $g^{\mathrm{real}}$ curve with respect to $x$ equals 0 corresponds to the nucleation/spinodal decomposition boundary (i.e., the spinodal curve) in the phase diagram. 
Mathematically, for the symmetric model, 
\begin{align}
\label{sym-binodal}
\left( \dfrac{\partial g^{\mathrm{real}}}{\partial x} \right)_{T, P} &= RT \ln \left( \dfrac{x}{1-x} \right) + (1-2x) w_\mathrm{sym}, \\
\label{sym-spinodal}
\left( \dfrac{\partial^2 g^{\mathrm{real}}}{\partial x^2} \right)_{T, P} &= \dfrac{RT}{x(1-x)} - 2w_\mathrm{sym}, 
\end{align}
and for the asymmetric model, 
\begin{align}
\label{asym-binodal}
\left( \dfrac{\partial g^{\mathrm{real}}}{\partial x} \right)_{T, P} &= RT \ln \left( \dfrac{x}{1-x} \right) + w_\mathrm{A} x (2-3x) + w_\mathrm{B} (3 x^2 - 4 x + 1), \\
\label{asym-spinodal}
\left( \dfrac{\partial^2 g^{\mathrm{real}}}{\partial x^2} \right)_{T, P} &= \dfrac{RT}{x(1-x)} + 2 \{ w_\mathrm{A} (1-3x) + w_\mathrm{B} (3x-2)\}, 
\end{align}
the solutions $x$ when these equations are equal to 0 form the binodal and spinodal curves. 
Binodal and spinodal curves appear on the cut surfaces of binodal and spinodal surfaces in temperature--pressure--chemical composition space (e.g., Aursand et al., 2017). Hence, either temperature or pressure can be selected for the vertical axis in the lower panel of Fig. \ref{Fig2}. 
This will be explained in section \ref{P-change}. 

It is well known that silicate melts, which are mixtures of multiple types of oxides, can separate into several distinct phases depending on changes in temperature and pressure, both in natural and industrial compositions: for example, liquid--liquid separation as seen in James (1975); Charlier and Grove (2012). 
Dehydration of magma is no exception and can be viewed as a separation into liquid phase (water-saturated silicate melt) and vapor phase (silicate-saturated water vapor), i.e., liquid--vapor separation. 
In this study, hydrous magma is approximated as a regular solution consisting of two components: anhydrous silicate (melt) and water (vapor), and the thermodynamics of mixing these two components is considered. Therefore, in the following, components A and B are taken to represent silicate and water, respectively. 
Note that water in magma exists as two molecular species: the molecule \ce{H2O}$_\mathrm{m}$ and the hydroxyl groups OH (e.g., Stolper, 1982a; 1982b), but we consider them together here. That is, $x$ is the mole fraction of total water. 
Notably, at and near the silicate end member, a crystalline phase precipitates at low temperatures (e.g., Fig. 4 in Ostrovsky, 1966 for the \ce{SiO2}--\ce{H2O} system; Fig. 7 in Paillat et al., 1992; and Fig. 3 in Makhluf et al., 2020 for the albite--\ce{H2O} system); however, in this study, it is neglected, assuming the amount to be minute and the system to be at a sufficiently high temperature for this assumption to hold, for simplicity.

\subsection{Regular solution approximation}
\subsubsection{Symmetric model vs. assymetric model} 
\label{bi-spi}
We will consider which model, symmetric or asymmetric, is more appropriate for approximating hydrous magma as a two-component regular solution in the silicate--water system, based on both observational and experimental facts about the behavior of real systems and the mathematical convenience of the models. First, we will discuss the relationship between $x_\mathrm{bi}$ and $x_\mathrm{spi}$ for each model. In particular, to later utilize knowledge on the silicate-rich side (e.g., the solubility of water in magma), we will focus on the range of $x$  smaller than the $x$ at which $g^\mathrm{real}$ reaches its maximum.

\begin{figure}
    \centering
    \includegraphics{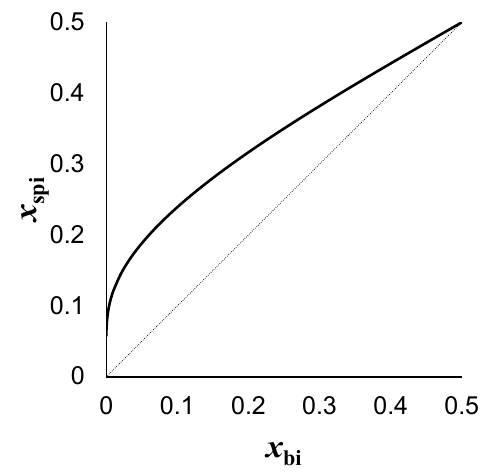}
    \caption{The relation between $x_\mathrm{bi}$ and $x_\mathrm{spi}$ at an arbitrary fixed temperature, derived from a series of Eqs. (\ref{sym-binodal}) and (\ref{sym-spinodal}), which represent the first- and second-order derivatives of $g^\mathrm{real}$ by $x$. $x_\mathrm{bi}$ and $x_\mathrm{spi}$ are the $x$ values that constitute the binodal and spinodal curves, respectively. The range $0 < x < 0.5$ corresponds to the left half of Fig. \ref{Fig1}. The relation $x_\mathrm{spi} > x_\mathrm{bi}$ holds for all the ranges.}
    \label{Fig3}
\end{figure}

For the symmetric model, combining Eqs. (\ref{sym-binodal}) and (\ref{sym-spinodal}) and eliminating $w_\mathrm{sym}$, the following relation between $x_\mathrm{bi}$ and $x_\mathrm{spi}$ is derived:
\begin{equation}
\label{bi-spi-relation}
x_\mathrm{spi} = \dfrac{1}{2} \left\{ 1 - \sqrt{1 - \dfrac{2 (1 - 2 x_\mathrm{bi})}{\ln \dfrac{1 - x_\mathrm{bi}}{ x_\mathrm{bi}}}} \right\}.
\end{equation}
This relation is shown in Fig. \ref{Fig3} in the $0 < x < 0.5$ range, corresponding to the left half area of Fig. \ref{Fig2}.

\begin{figure}
    \centering
    \includegraphics[width=1.1\linewidth]{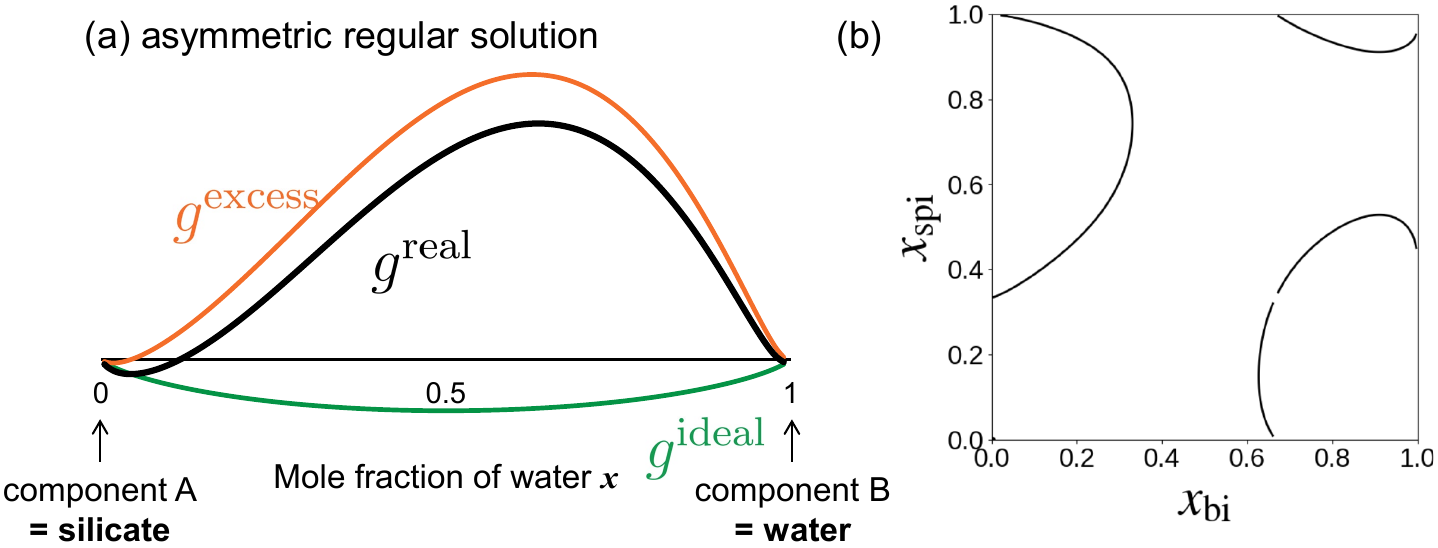} 
    \caption{(a) Schematic diagram of the molar Gibbs energy for a realistic hydrous magma, when it was assumed as an asymmetric regular solution of anhydrous silicate (melt) and water (vapor). The meaning of each curve is the same as in Fig. 1. Components A and B correspond to silicate and water, respectively. 
(b) The relationship between $x_\mathrm{bi}$ and $x_\mathrm{spi}$ in a two-component asymmetric regular solution model. An example is shown for the case where $T = 1,000^\circ$C, $w_\mathrm{A} = 345.7$ kJ/mol, and $w_\mathrm{B} = -1.0$ kJ/mol.}
    \label{Fig4}
\end{figure}

On the other hand, in the asymmetric model, the signs of the slopes at both ends of the $g^\mathrm{excess}$, i.e., $w_\mathrm{A}$ and $w_\mathrm{B}$, are important. When combining Eqs. (\ref{asym-binodal}) and (\ref{asym-spinodal}), if both $w_\mathrm{A}$ and $w_\mathrm{B}$ are positive, the relationship between $x_\mathrm{bi}$ and $x_\mathrm{spi}$ at a given $T$ or $P$ will be a one-to-one correspondence, just like in the symmetric model. As is well-known, the silicate--water system has a large miscibility gap (e.g., Kennedy, 1962; Paillat et al., 1992; Shen and Keppler, 1997; Bureau and Keppler, 1999), so it is necessary for $g^\mathrm{excess} > 0$ over a wide range of $x$. Therefore, the assumption that both $w_\mathrm{A} > 0$ and $w_\mathrm{B} > 0$ seems reasonable.
However, this case may not strictly align with the behavior of real systems. For example, at the silicate-rich side, when water dissolves, it is slightly exothermic, i.e., $g^\mathrm{excess} < 0$, as suggested by HF solution calorimetry of hydrous volcanic glasses synthesized at high temperature and pressure (Clemens and Navrotsky, 1987; Richet et al., 2004; 2006). If we trust this experimental result, the correct assumption would be that $w_\mathrm{A} > 0$ and slightly $w_\mathrm{B} < 0$.
In other words, the shape of the $g^\mathrm{excess}$ (and thus $g^\mathrm{real}$) for hydrous magma is asymmetric, and their peaks should slightly shift from $x = 0.5$ (Fig. \ref{Fig4} (a)). 
But then again, in this case, the relationship between $x_\mathrm{bi}$ and $x_\mathrm{spi}$ at a given $T$ or $P$ becomes a complex multivalued function, as shown in Fig. \ref{Fig4} (b). Specifically, this figure suggests that the spinodal curve corresponding to $x_\mathrm{bi}$ in the range $0 < x_\mathrm{bi} < 1/3$ intersects with the spinodal curve corresponding to $x_\mathrm{bi}$ in the range $2/3 < x_\mathrm{bi} < 1$, which is unrealistic.

Thus, the theoretical equations for a two-component asymmetric regular solution do not fit the actual energy curves of $g^\mathrm{excess}$ (and $g^\mathrm{real}$) well. This issue could be addressed by increasing the number of components to distinguish, thereby increasing the number of unknown interaction parameters $w$. For example, in the case of a three-component asymmetric regular solution, the expression for $g^\mathrm{excess}$ can be written as follows (Kakuda et al., 1994):
\begin{align}
g^\mathrm{excess}
&= x_\mathrm{A} x_\mathrm{A} x_\mathrm{B} w_\mathrm{AAB} \\ \notag
&+ x_\mathrm{A} x_\mathrm{B} x_\mathrm{B} w_\mathrm{ABB} \\ \notag
&+ x_\mathrm{B} x_\mathrm{B} x_\mathrm{C} w_\mathrm{BBC} \\ \notag
&+ x_\mathrm{B} x_\mathrm{C} x_\mathrm{C} w_\mathrm{BCC} \\ \notag
&+ x_\mathrm{C} x_\mathrm{C} x_\mathrm{A} w_\mathrm{CCA} \\ \notag
&+ x_\mathrm{C} x_\mathrm{A} x_\mathrm{A} w_\mathrm{CAA} \\ \notag
&+ 2 x_\mathrm{A} x_\mathrm{B} x_\mathrm{C} w_\mathrm{ABC},
\end{align}
where $w_\mathrm{ijk}$ is the interaction parameter between particles in a triplet consisting of three particles.
Nishiwaki (2023) distinguished between the molecular species of water (molecular water \ce{H2O_m} and hydroxyl groups \ce{OH}) and considered hydrous magma as a three-component system consisting of bridging oxygen of anhydrous silicate and these water species.
Additionally, he attempted to determine the values of the seven $w_\mathrm{ijk}$ parameters and the shapes of $g^\mathrm{excess}$ (and $g^\mathrm{real}$) over a wide range of temperatures and pressures. 
However, this problem is highly challenging from a linear algebraic standpoint, and the values reported in Nishiwaki (2023) are likely to be incorrect. 
We are currently working toward resolving this issue (Nishiwaki and Fukuya, in prep.).
In the end, although we know that the shape of $g^\mathrm{excess}$ (and $g^\mathrm{real}$) is complex, as shown in Fig. \ref{Fig4} (a), for now, we have no choice but to compromise and approximate hydrous magma as a two-component symmetric regular solution consisting of silicate and water. As a result, this approach is consistent with the scheme adopted by Allabar and Nowak (2018).

\subsubsection{Consistency with known phase diagrams} 
\label{assumptions}
Next, we will evaluate the validity of the two-component symmetric regular solution approximation by comparing it with the already-known phase diagrams.
In Fig. \ref{Fig2}, the regions on the silicate-rich and water-rich sides at pressures higher than the binodal curve, correspond to hydrous melt and aqueous fluid, respectively. Thus, when a pressure change occurs that cuts the binodal curve at a fixed chemical composition, the reaction ``magma (supercritical fluid) $\leftrightarrow$ water-saturated silicate melt (hydrous melt) + almost pure water vapor (aqueous fluid)'' occurs. The rightward reaction indicates exsolution with decompression, whereas the leftward reaction indicates mutual dissolution with compression. For example, according to the results of high-temperature and high-pressure experiments shown in Fig. 3 of Makhluf et al. (2020), for the albite--water system at 900$^{\circ}$C, the second critical endpoint (the vertex of the miscibility gap) is in the range of 1.25--1.40 GPa and 42--45 wt\%, i.e., $x =$ 0.57--0.60 (on a single oxygen basis). 
While some experimental studies have indicated that the position of the second critical endpoint may vary depending on the temperature and chemical composition of the silicate (Bureau and Keppler, 1999; Sowerby and Keppler, 2002), in this study, I assume that it does not deviate significantly from $x = 0.5$ to use the symmetric regular solution approximation, as mentioned in the previous section. 
This assumption may be somewhat forceful, but it is the simplest model we can present at this stage, and it will serve as a baseline for comparison when the detailed shape of the Gibbs energy is determined in the future, and the model is updated.

\subsection{Spinodal decomposition with pressure change}
\label{P-change}
Since spinodal decomposition is a phenomenon discovered in the field of inorganic materials such as ceramics and alloys (e.g., Cahn, 1965), it is typically discussed in terms of phase separation into solid--solid/solid--liquid/liquid--liquid systems that occur with decreasing temperature at normal pressure. 
Thus, temperature is typically used as the vertical axis when drawing phase diagrams. 
In contrast, when considering whether spinodal decomposition occurs in the phase separation of magma into gas--liquid systems, I assumed constant temperature and focused on the phase separation that occurs during decompression, since water solubility is much more dependent on pressure than temperature. Therefore, I adopted pressure as the vertical axis in the phase diagram in Fig. \ref{Fig2}. 
It has been suggested that, in nature, magma degassing may be more efficiently achieved through heating caused by the latent heat of crystallization, viscosity, and friction rather than decompression (e.g., Laval\'{e}e et al., 2015). However, typically, decompression experiments are conducted at a constant temperature, and so far, scenarios of spinodal decomposition have been proposed based solely on the results of such experiments (Gonnermann and Gardner, 2013; Allabar and Nowak, 2018; Allabar et al., 2020b; Sahagian and Carley, 2020; Gardner et al., 2023; Marks and Nowak, 2024; Hummel et al., 2024). Therefore, focusing on the pressure direction to test the validity of this scenario is not necessarily a flawed assumption. 

\section{Calculation methods} 
\label{methods}

\subsection{Relation between water solubility curve and binodal curve, and calculation of the spinodal curve}
\label{methods}
The section for $x < 0.5$ of the binodal curve on the $x$--$P$ plane at constant $T$, shown in the lower panel of Fig. \ref{Fig2}, should coincide with the solubility curve of water in the silicate melt at that temperature with respect to pressure change. In other words, the water solubility $c (P)$ in the silicate melt is equal to $x_\mathrm{bi} (P)$, which constitutes the binodal curve. From this and the relation Eq. (\ref{bi-spi-relation}) between $x_\mathrm{bi}$ and $x_\mathrm{spi}$, we can calculate the value of $x_\mathrm{spi}$ that constitutes the spinodal curve. The value of the silicate--water interaction parameter $w_\mathrm{sym} (P)$ at a fixed temperature can also be determined by substituting the value of $x_\mathrm{bi} (P)$ into Eq. (\ref{sym-binodal}) or the value of $x_\mathrm{spi} (P)$ into Eq. (\ref{sym-spinodal}).

\subsection{Conditions on the temperature, chemical composition, and water solubility in magma}
\label{Conditions}
Three types of silicate melts are assumed for temperature and chemical composition: K-phonolitic melt at 1,050$^{\circ}$C, basaltic melt at 1,100$^{\circ}$C, and albite melt at 900$^{\circ}$C. For the phonolitic melt, the conditions are the same as those used in all experiments of Allabar and Nowak (2018). Basaltic and albite melts were chosen to compare and examine the spinodal curves' behavior at higher pressures. The temperatures employed are those at which the pressure dependence of water solubility in the melt was already systematically determined from high-temperature and high-pressure experiments. According to Iacono-Marziano et al. (2007), who performed decompression experiments using AD79 Vesuvius pumice as did Allabar and Nowak (2018), since the water solubility in K-rich phonolitic melt at 1,050$^{\circ}$C agrees well with the value calculated from the empirical model of Moore et al. (1998), their formula was also used in this study in the range 0.1--300 MPa. Note that Moore et al. (1998) defined the mole fraction of water by treating each oxide (e.g., \ce{SiO2} and \ce{Al2O3}) as one unit, but this definition is no longer in common use, and here, the mole fraction was converted to the currently commonly used single-oxygen basis values (see Section 1.2 in Zhang, 1999). The water solubility in the basaltic melt at 1,100$^{\circ}$C was obtained by digitizing the fitting curve of the experimentally determined values for $\lesssim 600$ MPa, as shown in Fig. 2 of Hamilton et al. (1964). The water solubility in the albite melt at 900$^{\circ}$C was also obtained by digitizing the fitting curve of the experimentally determined values for $\lesssim 1000$ MPa, as shown in Fig. 8 of Burnham and Jahns (1962).

\section{Calculation results}
\label{results}
\begin{figure}
    \centering
    \includegraphics[width=1.15\linewidth]{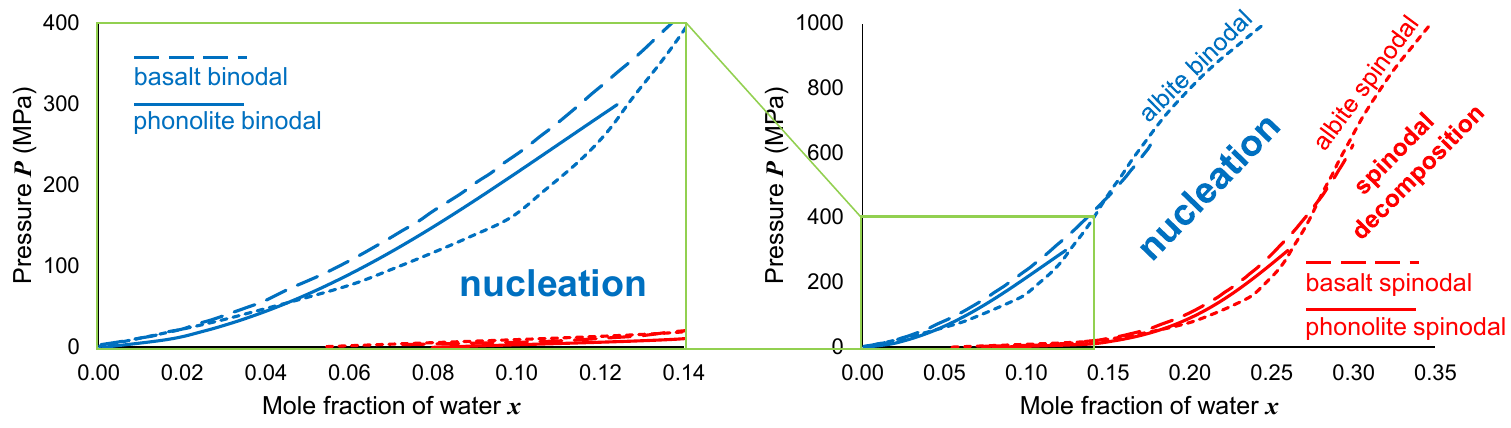}
    \caption{Binodal (blue) and spinodal (red) curves for hydrous 1,050$^{\circ}$C K-phonolitic (solid line), 1,100$^{\circ}$C basaltic (dashed line), and 900$^{\circ}$C albite (dotted line) melts in the pressure range 0.1--1000 MPa. The left panel shows an enlargement of the right panel at pressures below 400 MPa. The binodal curves correspond to the water solubility curves in the melt for each chemical composition (Moore et al., 1998; Hamilton et al., 1964; Burnham and Jahns, 1962). The position of spinodal curves was determined from the position of binodal curves and Eq. (\ref{bi-spi-relation}).}
    \label{Fig5}
\end{figure}

The binodal and spinodal curves drawn on the $x$--$P$ plane for the three silicate--water systems are shown in Fig. \ref{Fig5}. The geometric characteristics of both curves are similar, regardless of the silicate composition. Fig. 6 in Allabar and Nowak (2018) and Fig. 1 in Sahagian and Carley (2020) show a conceptual phase diagram in which both the binodal and spinodal curves are convex upward over the entire chemical composition range, with a large area inside the spinodal curve. The lower panel of Fig. \ref{Fig2} in this study is identical. However, when considering their actual position and shape quantitatively based on chemical thermodynamics, the solubility curve (= binodal curve) is convex downward at $\lesssim$ 400 MPa (approximately the relation $c (P)\propto P^{0.5}$ holds); therefore, the spinodal curve is also convex downward. In the pressure range examined in this study, the spinodal curve is situated at a much lower pressure relative to the binodal curve when fixed at a certain water content. For example, at $x = 0.10$, the water solubility (wt\%), $P_\mathrm{bi}$ (MPa), and $P_\mathrm{spi}$ (MPa) for the phonolitic melt, basaltic melt, and albite melt are (5.4, 219, $< 1$), (5.2, 239, 6), and (5.8, 166, 10), respectively. Furthermore, as $x$ increases, the corresponding $P_\mathrm{spi}$ gradually transitions to the high-pressure side, but the difference from $P_\mathrm{bi}$ is still very large. In other words, the nucleation region is much broader than the spinodal decomposition region, at least in the $x < 0.25$ range plotted in Fig. \ref{Fig5}. For example, in the case of albite melt, $P_\mathrm{bi} = 1000$ MPa, which corresponds to approximately $P_\mathrm{spi} = 190$ MPa.

\begin{figure}
    \centering
    \includegraphics{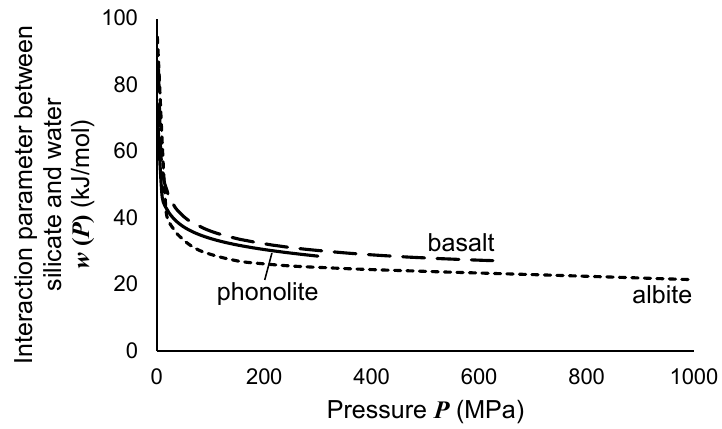}
    \caption{Calculation results of the interaction parameter between silicate and water $w (P)$ for hydrous K-phonolitic melt at 1,050$^{\circ}$C and 0.1--300 MPa (solid line), basaltic melt at 1,100$^{\circ}$C and 0.1--625 MPa (dashed line), and albite melt at 900$^{\circ}$C and 0.1--1000 MPa (dotted line).}
    \label{Fig6}
\end{figure}

The calculation results for $w (P)$ are shown in Fig. \ref{Fig6}. Although $w$ is large at 0.1 MPa (phonolite: 75 kJ/mol, basalt: 85 kJ/mol, and albite: 95 kJ/mol), for all three chemical compositions it monotonically decreases rapidly with increasing pressure over the entire pressure range. This behavior is consistent with the fact that the mutual dissolution of silicate and water proceeds at higher pressures at a fixed temperature, which narrows the miscibility gap. 

\section{Discussion}
\label{Discussion}
\subsection{Can spinodal decomposition occur during decompression-induced vesiculation of magma?}
\label{Can-spi}
First, I focus on the pressure range approximately below 300 MPa, where almost all decompression experiments have been conducted. 
All nucleation pressure values in 88 previous experiments, from Gardner et al. (1999) to Le Gall and Pichavant (2016), compiled and calculated by Shea (2017), fall within the nucleation region shown in Fig. \ref{Fig5} if variations in melt chemical composition are not considered. Additionally, all Allabar and Nowak (2018) experimental runs were performed in the range of the initial pressure of 200 MPa to final pressures of 70--110 MPa. Since the final pressures are higher than $P_\mathrm{spi}$, the nucleation pressure of these runs inevitably falls into the nucleation region. Therefore, the innovative scenario proposed by Allabar and Nowak (2018) and Sahagian and Carley (2020)---spinodal decomposition during decompression-induced vesiculation of magma---cannot occur in the pressure range of magmatic processes in the continental crust at depths of several hundred MPa. In addition, since $P_\mathrm{spi} << P_\mathrm{bi}$ as shown in Fig. \ref{Fig5}, for spinodal decomposition to occur without nucleation, it is necessary to maintain sufficient supersaturation, despite a very large decompression from an initial pressure higher than $P_\mathrm{bi}$ to a pressure lower than $P_\mathrm{spi}$. 
Gardner et al. (2023) stated that we must consider the overlap of various bubble formation mechanisms, including spinodal decomposition, to interpret BND and bubble size distributions in natural pyroclasts, I argue that we can focus only on homogeneous and heterogeneous nucleation as previously envisaged. 
Note that to confirm indeed that spinodal decomposition is unlikely to occur, it would likely be necessary to conduct runs with relatively rapid decompression to pressures lower than $P_\mathrm{spi}$. However, rapid decompression to near atmospheric pressure increases the possibility of capsule rupture, and bubbles' rapid expansion and coalescence may drastically overwrite the geometric arrangement of vesicular textures from its original state. 
Additionally, as mentioned in \ref{P-change}, in some natural systems where degassing is primarily driven by heating rather than decompression, the effective amount of decompression may become larger, making it possible that spinodal decomposition cannot be completely ruled out.

On the other hand, in regions of higher pressure and higher water content outside the drawn area of Fig. \ref{Fig5}, $P_\mathrm{spi}$ asymptotically approaches $P_\mathrm{bi}$, and spinodal decomposition is more likely to occur. 
In other words, spinodal decomposition may occur if decompression passes near the top of the second critical endpoint. 
In addition, because the second critical endpoint of silicate--water systems has been reported to shift to lower temperatures and pressures with increasing amounts of alkali metal oxides (e.g., \ce{Na2O} and \ce{K2O}) in the silicate (Bureau and Keppler, 1999; Sowerby and Keppler, 2002), Allabar and Nowak (2018) suggested that spinodal decomposition at low pressures may be more likely to occur in alkali-rich phonolite melts than in other silicic silicates. However, as long as the symmetric regular solution approximation is assumed, the shift of the second critical endpoint to lower pressures, i.e., the shift of the binodal curve to lower pressures, is accompanied by a shift of the spinodal curve to lower pressures because $x_\mathrm{bi}$ and $x_\mathrm{spi}$ change in tandem, as shown in Fig. \ref{Fig3}. 
In this case, the region of spinodal decomposition shown in Fig. \ref{Fig5} would be narrower, and spinodal decomposition would be less likely to occur. Therefore, if spinodal decomposition occurs in the phonolitic melt, the binodal curve of the phonolite--water system is expected to have a highly asymmetric shape to which the symmetric regular solution approximation cannot be applied. This might be related to the effective ionic radius of potassium (1.38 {\AA}) that is abundant in the phonolitic melt and is as large as that of oxygen (1.40 {\AA}) (Shannon, 1976). The packing ratio is higher than that in alkali-poor silicates (e.g., albite and rhyolite). Still, the shape of the binodal curve of the phonolite--water system has yet to be determined and requires exploration in detail using high-temperature and high-pressure experiments in the future. Considering additional complexities not accounted for in the symmetric model used in this study, it is difficult to rule out the possibility of spinodal decomposition completely.

\subsection{Consideration of kinetic effects} 
The discussions in the previous sections were based on equilibrium thermodynamics; thus, static binodal and spinodal curves were determined. 
However, in reality, during phase separation associated with pressure changes, kinetic effects arise due to the differences in the dynamic properties of silicate and water. As a consequence, the positions of the dynamic binodal and spinodal curves do not coincide with those of the static ones. 

Wang et al. (2021) conducted an in-situ observation of the phase separation process of peralkaline aluminosilicate (\ce{Na3AlSi5O13})--water system in a single-phase supercritical fluid near the second critical endpoint (approximately 700$^{\circ}$C and 1 GPa) using a hydrothermal diamond anvil cell. As a result, they observed that, at specific composition ratios (37--51 wt\% aluminosilicate), the network of hydrous silicate melt emerged as the temperature decreased, and two phases separated, with aqueous fluid enclosed within the silicate melt (spinodal decomposition). 
Additionally, Raman spectroscopy data collected in situ revealed that polymerized aluminosilicate species were selectively incorporated into the silicate melt, while silica monomers were selectively incorporated into the aqueous fluid. They interpreted this phenomenon by referring to the viscoelastic phase separation (VPS) theory for polymer solutions (Tanaka, 1994), suggesting that the relatively large molecules of the silicate polymer and the small molecules of water have significantly different relaxation timescales, i.e., viscoelastic properties, and thus, they undergo phase separation in a network-like structure.

Their experimental results using a peralkaline composition could offer a new perspective in interpreting the experimental results of Allabar and Nowak (2018), which also employed alkali-rich phonolite. Contrary to the traditional image of spinodal decomposition (the so-called ``intertwined structure'' shown in Fig. \ref{Fig2}), the experimental observation that phase separation occurs in a network-like structure suggests that the inference made from the experimental observations in Allabar and Nowak (2018)---where the spatial distribution of bubbles appeared to follow a specific wavelength and was attributed to spinodal decomposition---may not hold. 

\begin{figure}
    \centering
    \includegraphics[width=0.8\linewidth]{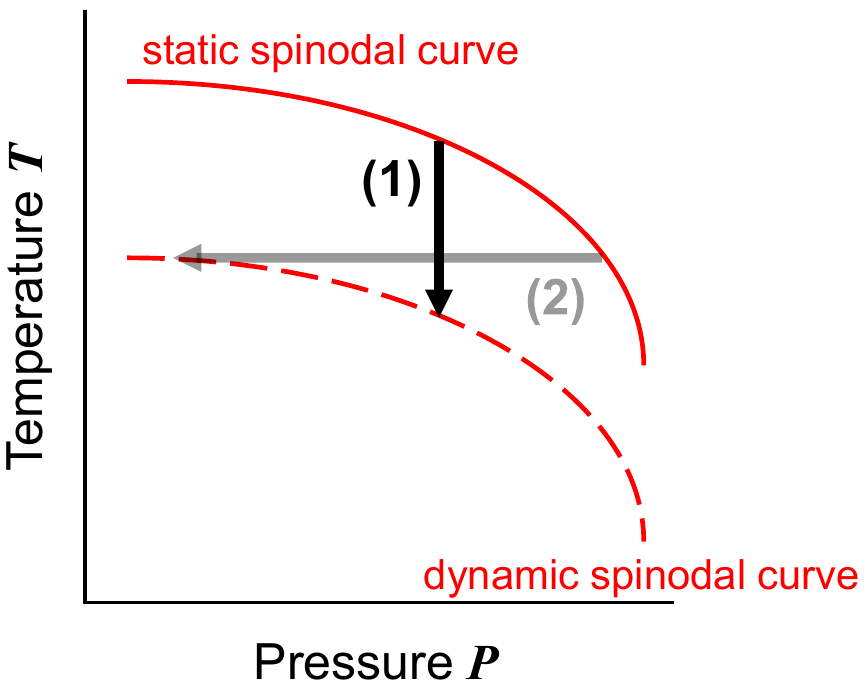}
    \caption{
The schematic relationship between the static and dynamic spinodal curves on the $T$--$P$ plane for a fixed composition. 
(1) At constant pressure, the dynamic spinodal curve shifts to the lower temperature side compared to the static one (Wang et al., 2021).
(2) In this case, at a fixed temperature, the spinodal curve shifts to the lower pressure side.}
    \label{Fig7}
\end{figure}

According to Wang et al. (2021), at constant pressure, the dynamic spinodal curve shifts to lower temperatures compared to the static one, as shown in (1) of Fig. \ref{Fig7}, where the spinodal curve was drawn with the negative slope, assuming that it would be similar to the negative slope of the phase boundary (binodal curve) appearing in the $T$--$P$ plane at a fixed composition. 
In this case, at a fixed temperature, the spinodal curve shifts to the lower pressure side: (2) in Fig. \ref{Fig7}. 
In other words, if the kinetic effects observed in their experiments were to appear in an isothermal decompression system, they would likely occur at pressures lower than the static spinodal pressure $P_\mathrm{spi}$ determined from thermodynamic calculations. However, as discussed in the previous section, within the pressure--composition (water content) range where decompression experiments have been conducted, $P_\mathrm{spi}$ is very low and lower than all the final pressures. Therefore, the scenario proposed by Allabar and Nowak (2018) would be even less likely to occur. 

\subsection{On the compilation of decompression-experimental results}
\label{manner}
Allabar and Nowak (2018) plotted their newly obtained experimental data and selected data from previous decompression experiments (Iacono-Marziano et al., 2007; Marxer et al., 2015; Preuss et al., 2016) conducted under the same chemical composition and physical conditions (phonolitic melt, 1,050$^{\circ}$C, initial pressure 200 MPa, and continuous decompression) on the decompression rate--BND plane (see Fig. 8 in their paper). 
The BND values of those data were all in the same order of magnitude, almost independent of the decompression rate. 
In other words, these data are non-harmonic with the classical nucleation theory's numerical prediction: BND $\propto |\mathrm{decompression\ rate}|^{1.5}$ (BND decompression rate meter from Toramaru, 2006). The authors considered that spinodal decomposition, rather than nucleation, might be occurring to explain these experimental results. 

The selection criteria for past data were not described in Allabar and Nowak (2018) but were detailed in Allabar et al. (2020a, b). 
In those studies, vesicle shrinkage during cooling and the initial water content dependence of BND in hydrous phonolitic melt were thoroughly investigated through precisely repeated experiments. They revealed that the several factors, denoted in \ref{AppB}, are essential in decompression experiments to determine BND.
Only the data obtained under these protocols (hereafter referred to as the ``good protocols'') are considered worth discussing in terms of consistency with the BND decompression meter (Allabar et al., 2020b). 
Though the good protocols were established through a series of experiments using phonolitic melt, since each is based on rational reasoning, they should be applied when conducting experiments with other compositions in the future.

\begin{figure}
    \centering
    \includegraphics[width=1.15\linewidth]{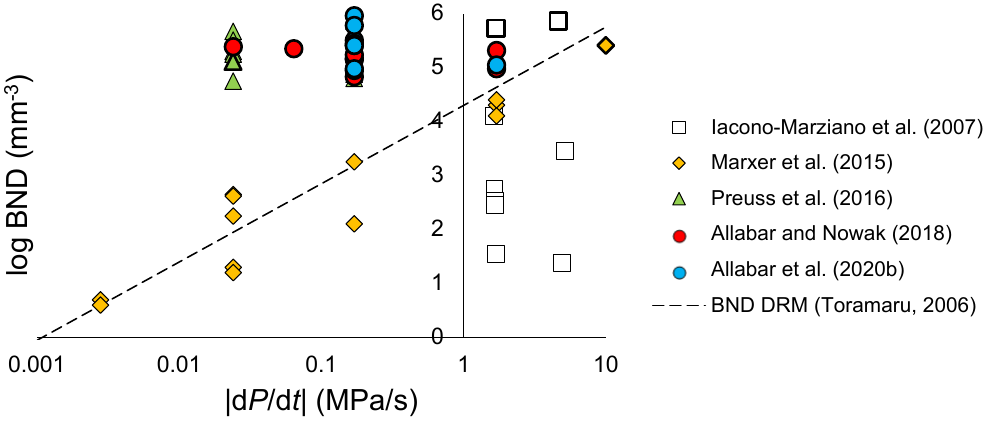}
    \caption{Results of decompression experiments using hydrous K-phonolitic melt from previous works (Iacono-Marziano et al., 2007; Marxer et al., 2015; Preuss et al., 2016; Allabar and Nowak, 2018; Allabar et al., 2020b). The dotted line is the theoretical prediction calculated by BND decompression rate meter, bubble number density decompression rate meter, from Toramaru (2006). The plots with thick borders represent reliable experimental data that meet the ``good protocols'' for decompression experiments (refer to the main text) described in Allabar et al. (2020b).
}
    \label{Fig8}
\end{figure}

I replotted all the experimental data on phonolitic melt at 1,050$^{\circ}$C, initial pressure 200 MPa, and continuous decompression in Fig. \ref{Fig8}. Here, in addition to the four papers mentioned earlier, data from Allabar et al. (2020b) Table 2 are referenced. The data that meet the good protocols are represented by plots with thick borders. 
As discussed in Allabar et al. (2020b), since BND has an initial water content dependence, there is some vertical variation even within populations of high BND values, but the variation is smaller compared to the data that mostly do not meet the good protocols from Iacono-Marziano et al. (2007) and Marxer et al. (2015). 
Even though Marxer et al. (2015)'s data does not satisfy the good protocols, it is harmonic with the BND decompression rate meter. Additionally, several previous papers reported harmonic results with the BND decompression rate meter (e.g., Mourtada-Bonnefoi and Laporte, 2004; Hamada et al., 2010), but they also seem not to satisfy the good protocols. 
In summary, there is a peculiar inconsistency where some data that do not satisfy the good protocols are in harmony with the BND decompression rate meter, while data that do satisfy the good protocols are not. 
I suspect that there are two main possible causes for this inconsistency. 
The first possibility, as mentioned in \ref{Can-spi}, is the uniqueness of the composition of phonolitic melt. 
It would be worth investigating how the BND decompression rate dependency behaves in the rhyolitic melt, which has relatively low alkali content and a wealth of past experimental examples when conducted following the good protocols. 
The second possibility, already discussed by Allabar et al. (2020b), is that the potential for heterogeneous nucleation on the surfaces of nanolites or ultrananolites such as Fe--Ti oxides (e.g., Mujin and Nakamura, 2014) cannot be excluded entirely. 
If the number density of such oxide crystals is a significant control factor for BND, then the BND obtained in experiments may not necessarily follow the BND decompression rate meter.

It might also be worth considering how to explain the independence of BND on the decompression rate without using spinodal decomposition---specifically, developing a new CNT-based theory, which could encompass this phenomenon, by improving the BND decompression rate meter. 
Assuming that all the experimental results plotted in Fig. \ref{Fig8} are equally reliable, they suggest that the decompression rate dependence of BND in phonolitic melts is extremely varied. 
It is known from magma crystallization experiments that the crystal number density can depend on the cooling rate (proportional to the 3/2 power of the cooling rate; consistent with the CNT-based prediction assuming diffusion-limited growth) as in Toramaru (2001) or shows no dependence, or even decreases (Martel and Schmidt, 2003; Cichy et al., 2011; Andrews and Befus, 2020). Toramaru and Kichise (2023) proposed that this wide range of cooling rate dependence can be explained by varying the pre-exponential factor of the nucleation rate and the surface tension. If there were a bubble version of this crystallization model, it might be possible to explain the wide variety of decompression rate dependence of BND (Toramaru, 2025).

\subsection{Application: Estimation of the surface tension between the melt and bubble nucleus}
When quantitatively considering the time evolution of the initial stage of magma vesiculation, values for several physical parameters (e.g., diffusion coefficient of water in the melt, viscosity of the melt, and surface tension between the melt and bubble nucleus) are required, as in the case of the BND decompression rate meter by Toramaru (2006). 
Many of these parameters have been carefully determined through past experiments. Still, even though the ``microscopic'' surface tension between the melt and (homogeneous spherical) bubble nucleus, $\sigma$, is a powerful parameter governing nucleation, its direct measurement is currently impossible. 
Therefore, it has been calculated only by fitting the integral value of the CNT-based nucleation rate $J$ over time with the obtained BND from decompression experiments: the inversion of BND using the CNT formula (e.g., Mourtada-Bonnefoi and Laporte, 2004; Cluzel et al., 2008; Hamada et al., 2010). 
However, since the detailed time evolution of $J$ is unknown, $\sigma$ has not been determined with high precision, and considerable uncertainty exists (Shea, 2017). 
See \ref{AppA} for the detailed expression of the equation of $J$ in CNT (e.g., Hirth et al., 1970). 
It has been found that $\sigma$ obtained via the CNT formula tends to be significantly smaller than the ``macroscopic'' surface tension $\sigma_\infty$ at the flat interface between the melt and vapor, directly measured by Bagdassarov et al. (2000) (e.g., Hamada et al., 2010). 

The reason why $\sigma < \sigma_\infty$ occurs was interpreted by Gonnermann and Gardner (2013) as follows.
According to the recent non-classical nucleation theory, the interface between the original and new phases loses sharpness and diffuses under non-equilibrium conditions (e.g., Chapter 4 in Kelton and Greer, 2010). 
Such interfacial diffusion should also occur during the vesiculation of magma with supersaturated water, which is no longer soluble due to decompression. 
In other words, the relationship $\sigma = \sigma_\infty$ (capillary approximation) holds at the binodal pressure $P_\mathrm{bi}$, but as the supersaturation increases, $\sigma$ is expected to decrease, and when the spinodal pressure $P_\mathrm{spi}$ is reached, where a distinct interface is expected to vanish, $\sigma$ should become zero.
They formulated this idea and derived $\sigma$ using a completely different approach from the BND inversion through CNT. 
However, at that time, $P_\mathrm{spi}$ was unknown, so it was considered a hypothetical parameter, and the uncertainty was incorporated into the mathematical treatment.
In this study, although a highly simplified equilibrium thermodynamic model is used, $P_\mathrm{spi}$ was determined for the first time. 
By substituting this into the equation for the dependence of $\sigma$ on the degree of supersaturation, the estimation of $\sigma$ is expected to be improved to a more straightforward way (Nishiwaki, in prep.).

\section{Conclusions}
\label{Conclusions}
I calculated the positions of the binodal and spinodal curves on the chemical composition--pressure plane by assuming that hydrous magma is a two-component symmetric regular solution of silicate and water and using the chemical thermodynamic equation and experimental data on water solubility in magma. The spinodal curve located significantly lower-pressure side than the binodal curve at pressures sufficiently below the second critical endpoint. The pressure ranges of all previous decompression experiments fell between these two curves. Therefore, decompression-induced vesiculation of magma occurs through nucleation, and spinodal decomposition is highly unlikely, in the magmatic processes associated with volcanic eruptions in most natural continental crusts. This result contradicts the recent inference that spinodal decomposition can occur based on observations of the bubble texture of decompression-experimental products. Additionally, by combining the spinodal pressure determined in this study with the non-classical nucleation theory, it may be possible to easily estimate the surface tension between the silicate melt and bubble nucleus, which has not been accurately determined. 

\appendix

\section{Equation of the nucleation rate in classical nucleation theory (CNT)}
\label{AppA}
The equation for $J$ in CNT (Hirth et al., 1970) is as follows:
\begin{equation}
\label{J}
J = \dfrac{2 n_0^2 D_\mathrm{\ce{H2O}} \overline{V}_\mathrm{\ce{H2O}}}{a_0} \sqrt{\dfrac{\sigma}{k_\mathrm{B} T}} \exp \left\{ - \dfrac{16 \pi \sigma^3}{3 k_\mathrm{B} T} (P^*_\mathrm{B} - P_\mathrm{M}) \right\},
\end{equation}
where $n_0$ is the number of water molecules per unit volume, $D_\mathrm{\ce{H2O}}$ is the diffusivity of total water in the melt, $\overline{V}_\mathrm{\ce{H2O}}$ is the partial molar volume of water in the melt, $a_0$ is the average distance between water molecules in the melt, $\sigma$ is the ``microscopic'' surface tension between the melt and the bubble nucleus, $k_\mathrm{B}$ is the Boltzmann's constant, $T$ is the temperature, $P^*_\mathrm{B}$ is the internal pressure of the critical bubble nucleus, and $P_\mathrm{M}$ is the pressure of the melt.

\section{``Good protocols'' established through a series of decompression experiments using phonolitic melt}
\label{AppB}
Based on the results of a series of decompression experiments conducted using phonolitic melt (Iacono-Marziano et al., 2017; Marxer et al., 2015; Preuss et al., 2016; Allabar and Nowak, 2018; Allabar et al., 2020b), Allabar et al. (2020b) proposed that ideal experimental products can be obtained by following the protocols outlined below:
(a) Homogeneous, bubble-free glass cylinders should be used as starting materials, encapsulated in noble metal tubes containing water for hydration at slightly \ce{H2O} undersaturated superliquidus conditions.
This setup prevents the presence of fluid bubbles before decompression.
(b) The sample should be continuously decompressed at a reasonable decompression time scale to mitigate the drainage of dissolved \ce{H2O} from the melt volume into heterogeneously formed vesicles at the capsule wall. This approach helps to prevent the potential detachment and movement of heterogeneously nucleated vesicles from the capsule walls into the melt volume.
(c) To determine the initial BND, the decompression should be terminated at a reasonable final pressure to avoid bubble coalescence which would reduce the initial BND.
(d) To determine bubble growth and the onset of bubble coalescence and to observe a possible second bubble formation event, a comprehensive set of experiments with small steps in final pressure down to low pressure is necessary.
(e) Subsequent quench of the partially degassed sample should be as fast as possible to minimize vesicle shrinkage. A cooling rate of $\sim$44$^{\circ}$C/s limits bubble shrinkage, inhibits quench crystal formation and avoids the formation of tension cracks.

\end{linenumbers}

\section*{Acknowledgments} 
The author would like to thank Marcus Nowak for sharing numerous detailed and insightful comments on the experiments and results from their team, which significantly contributed to the improvement of the manuscript.
The author is grateful to Yosuke Yoshimura for discussing spinodal decomposition in gas--liquid systems induced by pressure changes.
The author also appreciates Atsushi Toramaru for his insightful comments on the similarity to the latest crystallization models, which contributed to enhancing the content of this paper.
Special thanks go to Takeshi Ikeda for many discussions on the fundamentals of the thermodynamics of silicate--water immiscibility.
The reading circle for a glass science text with Shigeru Yamamoto inspired the author for this study.
Shumpei Yoshimura provided the author with some methods for calculating the mole fraction of water in magma.
The content of this paper was deepened after the discussion with Hidemi Ishibashi and Takayuki Nakatani.
The manuscript has been significantly improved through MP and Mathieu Colombier's peer review, as well as the generous handling by Editor Chiara Maria Petrone.
This work was supported by JSPS KAKENHI Grant Number JP23K19069.
The author would like to thank Editage (www.editage.jp) for English language editing.

\section*{References} 

Allabar, A., and Nowak, M., 2018. Message in a bottle: Spontaneous phase separation of hydrous Vesuvius melt even at low decompression rates. \textit{Earth Planet. Sci. Lett.} \textbf{501}, 192--201. \url{https://doi.org/10.1016/j.epsl.2018.08.047}


Allabar, A., Dobson, K. J., Bauer, C. C., and Nowak, M., 2020a. Vesicle shrinkage in hydrous phonolitic melt during cooling. \textit{Contrib. Mineral. Petrol.} \textbf{175} (3), 21. \url{https://doi.org/10.1007/s00410-020-1658-3}

Allabar, A., Salis Gross E., and Nowak, M., 2020b. The effect of initial \ce{H2O} concentration on decompression-induced phase separation and degassing of hydrous phonolitic melt. \textit{Contrib. Mineral. Petrol.} \textbf{175} (3), 22. \url{https://doi.org/10.1007/s00410-020-1659-2}

Andrews, B. J., and Befus, K. S., 2020. Supersaturation nucleation and growth of plagioclase: a numerical model of decompression-induced crystallization. \textit{Contrib. Mineral. Petrol.} \textbf{175} (3), 23. \url{https://doi.org/10.1007/s00410-020-1660-9}

Aursand, P., Gjennestad, M. A., Aursand, E., Hammer, M., and Wilhelmsen, \O., 2017. The spinodal of single- and multi-component fluids and its role in the development of modern equations of state. \textit{Fluid Phase Equilib.} \textbf{436}, 98--112. \url{https://doi.org/10.1016/j.fluid.2016.12.018}

Bagdassarov, N., Dorfman, A., and Dingwell, D. B., 2000. Effect of alkalis, phosphorus, and water on the surface tension of haplogranite melt. \textit{Am. Mineral.} \textbf{85} (1), 33--40. \url{https://doi.org/10.2138/am-2000-0105}

Bureau, H., and Keppler, H., 1999. Complete miscibility between silicate melts and hydrous fluids in the upper mantle: experimental evidence and geochemical implications. \textit{Earth Planet. Sci. Lett.} \textbf{165} (2), 187--196. \url{https://doi.org/10.1016/S0012-821X(98)00266-0}

Burnham, C. W., and Jahns, R. H., 1962. A method for determining the solubility of water in silicate melts. \textit{Am. J. Sci.} \textbf{260} (10), 721--745. \url{https://doi.org/10.2475/ajs.260.10.721}

Cahn, J. W., 1965. Phase separation by spinodal decomposition in isotropic systems. \textit{J. Chem. Phys.} \textbf{42} (1), 93--99. \url{https://doi.org/10.1063/1.1695731}

Cahn, J. W., and Hilliard, J. E., 1959. Free energy of a nonuniform system. III. Nucleation in a two-component incompressible fluid. \textit{J. Chem. Phys.} \textbf{31} (3), 688--699. \url{https://doi.org/10.1063/1.1730447}

Cassidy, M., Manga, M., Cashman, K., and Bachmann, O., 2018. Controls on explosive-effusive volcanic eruption styles. \textit{Nat. Commun.} \textbf{9} (1), 1--16. \url{https://doi.org/10.1038/s41467-018-05293-3}

Charlier, B., and Grove, T. L., 2012. Experiments on liquid immiscibility along tholeiitic liquid lines of descent. \textit{Contrib. Mineral. Petrol.} \textbf{164}, 27--44. \url{https://doi.org/10.1007/s00410-012-0723-y}

Cichy, S. B., Botcharnikov, R. E., Holtz, F., and Behrens, H., 2011. Vesiculation and microlite crystallization induced by decompression: a case study of the 1991--1995 Mt Unzen eruption (Japan). \textit{J. Petrol.} \textbf{52} (7--8), 1469--1492. \url{https://doi.org/10.1093/petrology/egq072}

Clemens, J. D., and Navrotsky, A., 1987. Mixing properties of \ce{NaAlSi3O8} melt--\ce{H2O}: new calorimetric data and some geological implications. \textit{J. Geol.} \textbf{95} (2), 173--188. \url{https://doi.org/10.1086/629118}

Cluzel, N., Laporte, D., Provost, A., and Kannewischer, I., 2008. Kinetics of heterogeneous bubble nucleation in rhyolitic melts: implications for the number density of bubbles in volcanic conduits and for pumice textures. \textit{Contrib. Mineral. Petrol.} \textbf{156} (6), 745--763. \url{https://doi.org/10.1007/s00410-008-0313-1}

Debenedetti, P. G., 2000. Phase separation by nucleation and by spinodal decomposition: Fundamentals. In: Kiran, E., Debenedetti, P. G., Peters, C. J. (Eds.), Supercritical Fluids. Nato Science Series \textbf{366}, 123--166. https://doi.org/10.1007/978-94-011-3929-8\verb|_|5

Gardner, J. E., Hilton, M., and Carroll, M. R., 1999. Experimental constraints on degassing of magma: isothermal bubble growth during continuous decompression from high pressure. \textit{Earth Planet. Sci. Lett.} \textbf{168} (1--2), 201--218. \url{https://doi.org/10.1016/S0012-821X(99)00051-5}


Gardner, J. E., Wadsworth, F. B., Carley, T. L., Llewellin, E. W., Kusumaatmaja, H., and Sahagian, D., 2023. Bubble formation in magma. \textit{Annu. Rev. Earth Planet. Sci.} \textbf{51}, 131--154. \url{https://doi.org/10.1146/annurev-earth-031621-080308}

Giachetti, T., Druitt, T. H., Burgisser, A., Arbaret, L., and Galven, C., 2010. Bubble nucleation, growth and coalescence during the 1997 Vulcanian explosions of Soufri\`{e}re Hills Volcano, Montserrat. \textit{J. Volcanol. Geotherm. Res.} \textbf{193} (3--4), 215--231. \url{https://doi.org/10.1016/j.jvolgeores.2010.04.001}

Gonnermann, H. M., and Gardner, J. E., 2013. Homogeneous bubble nucleation in rhyolitic melt: Experiments and nonclassical theory. \textit{Geochem. Geophys. Geosyst.} \textbf{14} (11), 4758--4773. \url{https://doi.org/10.1002/ggge.20281}

Guggenheim, E. A., 1952. Mixtures: the theory of the equilibrium properties of some simple classes of mixtures solutions and alloys. Clarendon Press. 

Haasen, P., 1996. Physical metallurgy. Cambridge university press.




Hamilton, D. L., Burnham, C. W., and Osborn, E. F., 1964. The solubility of water and effects of oxygen fugacity and water content on crystallization in mafic magmas. \textit{J. Petrol.} \textbf{5} (1), 21--39. \url{https://doi.org/10.1093/petrology/5.1.21}

Hirth, J. P., Pound, G. M., and St Pierre, G. R., 1970. Bubble nucleation. \textit{Metall. Trans.} \textbf{1} (4), 939--945. \url{https://doi.org/10.1007/BF02811776}

Houghton, B. F., Carey, R. J., Cashman, K. V., Wilson, C. J., Hobden, B. J., and Hammer, J. E., 2010. Diverse patterns of ascent, degassing, and eruption of rhyolite magma during the 1.8 ka Taupo eruption, New Zealand: evidence from clast vesicularity. \textit{J. Volcanol. Geotherm. Res.} \textbf{195} (1), 31--47. \url{https://doi.org/10.1016/j.jvolgeores.2010.06.002}

Hummel, F., Marks, P. L., and Nowak, M., 2024. Preparatory experiments to investigate the vesicle formation of hydrous lower Laacher See phonolite at near liquidus conditions. EGU General Assembly 2024, Vienna, Austria, EGU24-19758. \url{https://doi.org/10.5194/egusphere-egu24-19758}

Iacono-Marziano, G., Schmidt, B. C., and Dolfi, D., 2007. Equilibrium and disequilibrium degassing of a phonolitic melt (Vesuvius AD 79 ``white pumice'') simulated by decompression experiments. \textit{J. Volcanol. Geotherm. Res.} \textbf{161} (3), 151--164. \url{https://doi.org/10.1016/j.jvolgeores.2006.12.001}

James, P. F., 1975. Liquid-phase separation in glass-forming systems. \textit{J. Mater. Sci.} \textbf{10}, 1802--1825.\url{https://doi.org/10.1007/BF00554944}


Kakuda, Y., Uchida, E., and Imai, N., 1994. A new model of the excess Gibbs energy of mixing for a regular solution. \textit{Proc. Jpn. Acad. B} \textbf{70} (10), 163--168. \url{https://doi.org/10.2183/pjab.70.163}


Kennedy, G. C., 1962. The upper three-phase region in the system \ce{SiO2}--\ce{H2O}. \textit{Am. J. Sci.} \textbf{260}, 501--521. \url{https://doi.org/10.2475/AJS.260.7.501}

Lavall\'{e}e, Y., Dingwell, D. B., Johnson, J. B., Cimarelli, C., Hornby, A. J., Kendrick, J. E., von Aulock, F. W., Kennedy, B. M., Andrews, B. J., Wadsworth, F. B., Rhodes, E., and Chigna, G., 2015. Thermal vesiculation during volcanic eruptions. \textit{Nature} \textbf{528} (7583), 544--547. \url{https://doi.org/10.1038/nature16153}

Le Gall, N., and Pichavant, M., 2016. Homogeneous bubble nucleation in \ce{H2O}- and \ce{H2O}--{CO2}- bearing basaltic melts: Results of high temperature decompression experiments. \textit{J. Volcanol. Geotherm. Res.} \textbf{327}, 604--621. \url{https://doi.org/10.1016/j.jvolgeores.2016.10.004}

Makhluf, A. R., Newton, R. C. and Manning, C. E., 2020. Experimental investigation of phase relations in the system \ce{NaAlSi3O8}--\ce{H2O} at high temperatures and pressures: Liquidus relations, liquid-vapor mixing, and critical phenomena at deep crust-upper mantle conditions. \textit{Contrib. Mineral. Petrol.} \textbf{175} (8), 76. \url{https://doi.org/10.1007/s00410-020-01711-2}

Marks, P. L., and Nowak M., 2024. Decoding the \ce{H2O} phase separation mechanism as the trigger for the explosive eruption of the Lower Laacher See phonolite. EGU General Assembly 2024, Vienna, Austria, EGU24-7723. \url{https://doi.org/10.5194/egusphere-egu24-7723}

Martel, C., and Schmidt, B. C., 2003. Decompression experiments as an insight into ascent rates of silicic magmas. \textit{Contrib. Mineral. Petrol.} \textbf{144} (4), 397--415. \url{https://doi.org/10.1007/s00410-002-0404-3}

Marxer, H., Bellucci, P., and Nowak, M., 2015. Degassing of \ce{H2O} in a phonolitic melt: A closer look at decompression experiments. \textit{J. Volcanol. Geotherm. Res.} \textbf{297}, 109--124. \url{https://doi.org/10.1016/j.jvolgeores.2014.11.017}

Moore, G., Vennemann, T., and Carmichael, I. S. E., 1998. An empirical model for the solubility of \ce{H2O} in magmas to 3 kilobars. \textit{Am. Mineral.} \textbf{83} (1--2), 36--42. \url{https://doi.org/10.2138/am-1998-1-203}


Mujin, M., and Nakamura, M., 2014. A nanolite record of eruption style transition. \textit{Geology} \textbf{42} (7), 611--614. \url{https://doi.org/10.1130/G35553.1}

Murase, T., and McBirney, A. R., 1973. Properties of some common igneous rocks and their melts at high temperatures. \textit{Geol. Soc. Am. Bull.} \textbf{84} (11), 3563--3592. \url{https://doi.org/10.1130/0016-7606(1973)84<3563:POSCIR>2.0.CO;2}


Nguyen, C. T., Gonnermann, H. M., and Houghton, B. F., 2014. Explosive to effusive transition during the largest volcanic eruption of the 20th century (Novarupta 1912, Alaska). \textit{Geology} \textbf{42} (8), 703--706. \url{https://doi.org/10.1130/G35593.1}

Nishiwaki, M., 2023. Chemical-thermodynamic explorations on the dissolution of water in magma: Breaking of the ideal mixing model and estimations of temperature change with decompression-induced vesiculation. Doctoral Dissertation, Kyushu University. \url{https://catalog.lib.kyushu-u.ac.jp/opac_detail_md/?reqCode=frombib&lang=0&amode=MD823&opkey=B168673756684664&bibid=6787423&start=1&bbinfo_disp=0}



Paillat, O., Elphick, S. C., and Brown, W. L., 1992. The solubility of water in \ce{NaAlSi3O8} melts: a re-examination of Ab--\ce{H2O} phase relationships and critical behaviour at high pressures. \textit{Contrib. Mineral. Petrol.} \textbf{112}, 490--500. \url{https://doi.org/10.1007/BF00310780}

Preuss, O., Marxer, H., Ulmer, S., Wolf, J., and Nowak, M., 2016. Degassing of hydrous trachytic Campi Flegrei and phonolitic Vesuvius melts: Experimental limitations and chances to study homogeneous bubble nucleation. \textit{Am. Mineral.} \textbf{101} (4), 859--875. \url{https://doi.org/10.2138/am-2016-5480}

Richet, P., Hovis, G., and Whittington, A., 2006. Water and magmas: Thermal effects of exsolution. \textit{Earth Planet. Sci. Lett.} \textbf{241} (3--4), 972--977. \url{https://doi.org/10.1016/j.epsl.2005.10.015}

Richet, P., Hovis, G., Whittington, A., and Roux, J., 2004. Energetics of water dissolution in trachyte glasses and liquids. \textit{Geochim. Cosmochim. Acta} \textbf{68} (24), 5151--5158. \url{https://doi.org/10.1016/j.gca.2004.05.050}

Sahagian, D., and Carley, T. L., 2020. Explosive volcanic eruptions and spinodal decomposition: A different approach to deciphering the tiny bubble paradox. \textit{Geochem. Geophys. Geosyst.} \textbf{21} (6), e2019GC008898. \url{https://doi.org/10.1029/2019GC008898}


Shannon, R. D., 1976. Revised effective ionic radii and systematic studies of interatomic distances in halides and chalcogenides. \textit{Acta Crystallogr. A: Found. Adv.} \textbf{32} (5), 751--767. \url{https://doi.org/10.1107/S0567739476001551}

Shea, T., 2017. Bubble nucleation in magmas: A dominantly heterogeneous process?. \textit{J. Volcanol. Geotherm. Res.} \textbf{343}, 155--170. \url{https://doi.org/10.1016/j.jvolgeores.2017.06.025}

Shen, A. H., and Keppler, H., 1997. Direct observation of complete miscibility in the albite--\ce{H2O} system. Nature \textbf{385} (6618), 710--712. \url{https://doi.org/10.1038/385710a0}

Shimozuru, D., Nakamuda, O., Seno, H., Noda, H., and Taneda, S., 1957. Mechanism of pumice formation. Bull. Volcanol. Soc. Japan \textbf{2}, 17--25. 

Sowerby, J. R., and Keppler, H., 2002. The effect of fluorine, boron and excess sodium on the critical curve in the albite--\ce{H2O} system. \textit{Contrib. Mineral. Petrol.} \textbf{143} (1), 32--37. \url{https://doi.org/10.1007/s00410-001-0334-5}

Sparks, R. S. J., 1978. The dynamics of bubble formation and growth in magmas: A review and analysis. \textit{J. Volcanol. Geotherm. Res.} \textbf{3} (1--2), 1--37. \url{https://doi.org/10.1016/0377-0273(78)90002-1}

Stolper, E., 1982a. Water in silicate glasses: An infrared spectroscopic study. \textit{Contrib. Mineral. Petrol.} \textbf{81} (1), 1--7. \url{https://doi.org/10.1007/BF00371154}

Stolper, E., 1982b. The speciation of water in silicate melts. \textit{Geochim. Cosmochim. Acta} \textbf{46} (12), 2609--2620. \url{https://doi.org/10.1016/0016-7037(82)90381-7}

Tanaka, H., 1994. Critical dynamics and phase‐separation kinetics in dynamically asymmetric binary fluids: New dynamic universality class for polymer mixtures or dynamic crossover?. \textit{J. Chem. Phys.} \textbf{100} (7), 5323--5337. \url{https://doi.org/10.1063/1.467197}



Toramaru, A., 1989. Vesiculation process and bubble size distributions in ascending magmas with constant velocities. \textit{J. Geophys. Res. Solid Earth} \textbf{94} (B12), 17523--17542. \url{https://doi.org/10.1029/JB094iB12p17523}

Toramaru, A., 1995. Numerical study of nucleation and growth of bubbles in viscous magmas. \textit{J. Geophys. Res. Solid Earth} \textbf{100} (B2), 1913--1931. \url{https://doi.org/10.1029/94JB02775}

Toramaru, A., 2001. A numerical experiment of crystallization for a binary eutectic system with application to igneous textures. \textit{J. Geophys. Res. Solid Earth} \textbf{106} (B3), 4037--4060. \url{https://doi.org/10.1029/2000JB900367}

Toramaru, A., 2006. BND (bubble number density) decompression rate meter for explosive volcanic eruptions. \textit{J. Volcanol. Geotherm. Res.} \textbf{154} (3--4), 303--316. \url{https://doi.org/10.1016/j.jvolgeores.2006.03.027}

Toramaru, A., 2022. Vesiculation and crystallization of magma: Fundamentals of volcanic eruption process, conditions for magma vesiculation. Springer Singapore. \url{https://doi.org/10.1007/978-981-16-4209-8}

Toramaru, A., 2025. The theoretical basis for textural indices of eruption dynamics: review and new conceptual models. \textit{Earth Planets Space} \textbf{77}, 27. 
\url{https://doi.org/10.1186/s40623-025-02146-4}

Toramaru, A., and Kichise, T., 2023. A new model of crystallization in magmas: Impact of pre-exponential factor of crystal nucleation rate on cooling rate exponent and log-linear crystal size distribution. \textit{J. Geophys. Res. Solid Earth} \textbf{128} (10), e2023JB026481. \url{https://doi.org/10.1029/2023JB026481}

Verhoogen, J., 1951. Mechanism of ash formation. \textit{Am. J. Sci.} \textbf{249}, 723--739.


Wang, Q. X., Zhou, D. Y., Li, W. C., and Ni, H. W., 2021. Spinodal decomposition of supercritical fluid forms melt network in a silicate--\ce{H2O} system. \textit{Geochem. Perspect. Lett.} \textbf{18}, 22--26. \url{https://doi.org/10.7185/geochemlet.2119}

Zhang, Y., 1999. \ce{H2O} in rhyolitic glasses and melts: measurement, speciation, solubility, and diffusion. \textit{Rev. Geophys.} \textbf{37} (4), 493--516. \url{https://doi.org/10.1029/1999RG900012}

\section*{CRediT authorship contribution statement}
Mizuki Nishiwaki:  Conceptualization, Methodology, Formal analysis, Investigation, Writing -- Original Draft, Writing -- Review \& Editing, Visualization, Project administration, Funding acquisition.

\section*{Declaration of competing interest}
The author declares that he has no known competing financial interests or personal relationships that could have appeared to influence the work reported in this paper.

\section*{Data availability}
The author confirms that the data supporting the findings of this study are available within the article.

\end{document}